\documentclass[runningheads, breaklinks=true]{llncs}
\usepackage{marvosym}
\usepackage{orcidlink}
\usepackage[T1]{fontenc}
\usepackage{graphicx}
\usepackage{color}
\usepackage{amsmath}
\usepackage{amssymb}
\usepackage{amsfonts}
\usepackage{booktabs}
\usepackage{makecell}
\usepackage{multirow}
\usepackage{enumitem}
\usepackage{dsfont}
\usepackage{float}
\usepackage{flafter}
\usepackage{placeins}
\usepackage{microtype}
\usepackage{xurl}

\usepackage{eso-pic}

\begin{document}

 \title{EI-DDLGN: Efficient Encrypted Inference with Deep Differentiable Logic Gate Networks under TFHE}

\titlerunning{EI-DDLGN: Efficient Encrypted Inference under TFHE}
\author{Mahmoud Y. M. Yassin\inst{1}\orcidlink{0009-0006-1405-8379}(\textrm{\Letter}) \and
Mahmoud AbdelHafeez Sayed\inst{1} \and
Mostafa Taha\inst{1}}

\authorrunning{M. Y. M. Yassin et al.}

\institute{Systems and Computer Engineering, Carleton University,
Ottawa, ON, Canada\\
\email{MahmoudYassin3@cmail.carleton.ca}\\
\email{\{Mahmoud.Sayed3, Mostafa.Taha\}@Carleton.ca}}

\maketitle

\AddToShipoutPictureBG*{%
  \AtTextUpperLeft{\raisebox{2.5\baselineskip}{%
    \parbox{\textwidth}{\centering\footnotesize\itshape
      Accepted at the 19th International Symposium on Foundations
      \& Practice of Security (FPS 2026)}}}%
}

\begin{abstract}
Privacy-preserving inference via Torus Fully Homomorphic Encryption (TFHE) provides strong protection for sensitive data in outsourced deep learning applications. However, most TFHE-compatible neural network frameworks remain based on arithmetic neural architectures, resulting in high inference latency due to programmable bootstrapping (PBS), accumulator growth, and circuit bit-width sensitivity. In this work, we investigate Deep Differentiable Logic Gate Networks (DDLGNs) as a Boolean-native alternative for encrypted inference under TFHE. Because DDLGNs learn Boolean computations directly and discretize into fixed logic gate networks, their inference procedure is naturally aligned with TFHE's Boolean execution model and avoids arithmetic accumulation in hidden layers. We present EI-DDLGN, the first in-depth study of TFHE-based DDLGN inference, and characterize how encrypted execution cost depends on model size, learned Boolean-function distribution, and propagated wire status. We also introduce Model-Fixed-Wire PBS Bypass (MFW-PBS Bypass), a semantics-preserving execution strategy that eliminates unnecessary PBS operations without modifying the learned network topology. Evaluations across 72 depth-width configurations on MNIST, FashionMNIST, and UCI Phishing show that DDLGNs constitute an efficient alternative to arithmetic TFHE inference, achieving substantially improved accuracy--latency trade-offs. Notably, on MNIST, EI-DDLGN-Small matches the accuracy of QAT-FCNN-4 while reducing encrypted inference latency by \(13.4\times\). Our implementation is available at \url{https://github.com/Carleton-SCI/EI-DDLGN}.
\end{abstract}

\section{Introduction}

Privacy-preserving inference enables a server to evaluate a deep learning model without accessing a client's plaintext input. Among the available approaches, Torus Fully Homomorphic Encryption (TFHE) is particularly attractive because it supports arbitrary computations on encrypted data and provides efficient programmable bootstrapping (PBS) for evaluating nonlinear functions \cite{chillotti2020tfhe,chillotti2021programmable}.

Recent TFHE-based inference frameworks have primarily focused on arithmetic neural networks adapted through Quantization-Aware Training (QAT) \cite{stoian2023deep}. Although quantization reduces encrypted computation cost, inference latency remains strongly influenced by accumulator growth, circuit bit-width, and PBS requirements. Binary Neural Networks (BNNs) further reduce arithmetic complexity by restricting weights and activations to binary values \cite{huang2025privacy}. However, they remain weight-based architectures whose encrypted execution still relies on accumulation operations.

Deep Differentiable Logic Gate Networks (DDLGNs) \cite{petersen2022deep} provide a fundamentally different approach. Instead of representing a neuron as a weighted arithmetic operator, DDLGNs learn Boolean computations directly using differentiable logic gates that are discretized after training into a fixed logic gate network. Consequently, inference is expressed entirely as Boolean computation and is naturally aligned with TFHE's Boolean execution model.

This architectural alignment has not been studied previously. In particular, it remains unclear how learned Boolean function distributions affect encrypted execution cost, how TFHE bootstrapping requirements emerge from the resulting logic networks, and whether properties of the learned logic structure can be exploited to further reduce encrypted evaluation cost.

To address these questions, we present EI-DDLGN, the first in-depth study of DDLGN inference under TFHE. We characterize the PBS cost structure of TFHE-based DDLGN inference, showing that encrypted execution depends jointly on model size, learned Boolean-function distribution, and propagated wire status. We further introduce Model-Fixed-Wire PBS Bypass (MFW-PBS Bypass), a semantics-preserving execution method that eliminates unnecessary PBS operations without modifying the learned network topology.

We evaluate 72 DDLGN configurations across MNIST, FashionMNIST, and UCI Phishing datasets. The results demonstrate clear accuracy--latency and accuracy--PBS budget trade-offs and show that DDLGNs constitute an efficient alternative to arithmetic TFHE inference. Notably, EI-DDLGN-Small matches the accuracy of QAT-FCNN-4 \cite{stoian2023deep} while reducing encrypted inference latency by \(13.4\times\).

\medskip

\noindent\textbf{Contributions:}

\begin{itemize}
\item We present EI-DDLGN, the first in-depth study of DDLGN inference under TFHE and demonstrate that Boolean-native neural architectures constitute an efficient alternative to arithmetic TFHE inference.

\item We characterize the encrypted cost structure of TFHE-based DDLGN inference and show that it depends jointly on model size, learned Boolean-function distribution, and propagated wire status.

\item We introduce Model-Fixed-Wire PBS Bypass (MFW-PBS Bypass), a semantics-preserving execution method that bypasses unnecessary PBS operations without modifying the learned network topology.

\item We evaluate 72 models across three datasets, analyze accuracy--latency and accuracy--PBS budget trade-offs, and demonstrate up to a \(13.4\times\) latency reduction over an accuracy-matched QAT-FCNN baseline.
\end{itemize}

\section{Related Work}
\label{Sec:Related Work}
Privacy-preserving neural network inference under homomorphic encryption has been 
studied through several architectural and cryptographic design choices. 
Early and recent TFHE-based works showed that PBS can refresh ciphertext noise while evaluating nonlinear functions through lookup tables, which enables arbitrary-depth encrypted inference but still leaves PBS latency as the main bottleneck ~\cite{chillotti2021programmable}. 
For example, FHE--DiNN uses post-training discretized feedforward networks under TFHE, where at inference, weighted sums are evaluated homomorphically, and the sign activation is handled during bootstrapping \cite{bourse2018fast}. More recent QAT-FCNN frameworks quantize weights and activations during training so that inference can be expressed over low-bit integer representations ~\cite{stoian2023deep}. 
This keeps the network close to the arithmetic NN formulation, but the encrypted cost remains sensitive to accumulator growth, circuit bit-width, pruning, and the number of PBS operations.

Binary neural networks (BNNs) also reduce arithmetic cost by using binary weights and activations, 
and recent TFHE-oriented variants use operational logic and fewer bootstrapping operations to improve encrypted inference efficiency ~\cite{huang2025privacy}. 
However, BNNs are still weight-based neural networks. Their Boolean operations are mainly used to approximate or accelerate arithmetic computation such as dot 
products. This is different from Deep Differentiable Logic Gate Networks (DDLGNs), introduced by Petersen \emph{et al.} ~\cite{petersen2022deep}. 
In DDLGNs, the neuron itself is redefined as a learned two-input logic operation. After training through a differentiable relaxation, 
the model is discretized into a fixed logic gate network. Thus, DDLGNs do not merely approximate arithmetic NNs with binary operations;  they learn the Boolean computation directly.

From a TFHE perspective, existing approaches can be broadly divided into two categories. The first category preserves the arithmetic neural network abstraction and attempts to reduce its encrypted cost through quantization, pruning, or binarization. QAT-FCNNs and TFHE-oriented BNNs belong to this family. The second category replaces the arithmetic neuron itself with a Boolean computation primitive. DDLGNs belong to this category and therefore provide an opportunity to study encrypted inference from a fundamentally different computational perspective.

\begin{table}[htbp]
\centering
\caption{Comparison between QAT-FCNN \cite{stoian2023deep}, BNNs \cite{huang2025privacy}, and DDLGNs \cite{petersen2022deep} for encrypted inference.}
\label{tab:model_family_comparison}
\scriptsize
\setlength{\tabcolsep}{3pt}
\begin{tabular}{p{0.17\linewidth}p{0.40\linewidth}p{0.34\linewidth}}
\toprule
Model family & What the neuron represents & Main TFHE implication \\
\midrule
QAT-FCNN & Quantized weighted neuron; the arithmetic NN is preserved after QAT & Cost depends on circuit bit-width, accumulator growth, and PBS count \\
BNNs & Binary weighted neuron; Boolean logic mainly approximates arithmetic/dot-product computation but still relies on accumulation operations & Still weight-based; requires deep adder trees and high bootstrapping cost. \\
DDLGNs & Learned two-input logic gate; the perceptron is redefined as one learned logic operation & Boolean-native network; cost follows learned functions and executed PBS count \\
\bottomrule
\end{tabular}
\end{table}

The distinction summarized in Table~\ref{tab:model_family_comparison} is the main reason behind our proposed study of EI-DDLGN. 
Rather than adapting arithmetic NNs to TFHE through quantization or binarization, EI-DDLGN starts from a Boolean-native model whose inference is already expressed as learned Boolean operations. Consequently, the dominant factors governing encrypted execution differ from those of arithmetic TFHE inference and are more directly related to the learned logic structure itself. 

Privacy-preserving inference has also been studied for CNNs under homomorphic encryption, including CryptoNets, SHE, DCT-domain inference, model redesign for FHEW/TFHE, Truth-Table Convolutions (TT‑TFHE), and hybrid encrypted lookup-table frameworks \cite{gilad2016cryptonets,lou2019she,roy2024dct,ku2025optimizing,benamira2025tttfhe,li2026fsat,li2025cryptdnn}. 
These works show that encrypted inference can scale to richer vision models, but convolutional layers introduce a different set of encrypted computation challenges. 
Accordingly, this work focuses on feedforward DDLGNs as the cleanest setting for exposing the learned function cost structure, PBS behavior, and bypass opportunities of Boolean-native TFHE inference, while leaving convolution-specific challenges to future work.

\section{Preliminaries}
\label{Sec: Preliminaries}
This section briefly reviews the cryptographic and model background needed for the remainder of the paper. We first summarize the main principles of TFHE, then introduce DDLGNs and explain how their Boolean-native structure differs from both QAT-based NNs and BNNs, and what makes them particularly suitable for encrypted inference.
\subsection{TFHE}
 TFHE encrypts a message by embedding it into a noisy ciphertext, based essentially on the learning with error (LWE) problem for security. Let $m$ denote a plaintext message and let $e$ denote a small error term. A ciphertext can be viewed as
\begin{equation} \label{eq:enc_1}
    \begin{split}
        &\mathbf{c} \leftarrow \text{LWE}_{\mathbf{s}}(m) = (a_1,a_2, ..., a_n, b)\in \mathds{Z}_q^{n+1},\\
        &b = \sum_{i=1}^{n}a_i\cdot s_i + e + m,
    \end{split}
\end{equation}
where $n$ is the LWE dimension, $q$ is the ciphertext modulus, $\mathbf{a}=(a_1,...,a_n)\overset{\$}{\leftarrow} \mathds{Z}_q$ represents a uniformly random vector and $\mathbf{s}\in\{0,1\}^n$ is the binary secret key. The construction of~(\ref{eq:enc_1}) implies that linear operations between ciphertext experience direct homomorphism with the underlying plaintext values; however, non-linear operations need a different approach. Formally speaking, a complete operation between two ciphertexts $\mathbf{c}_1$ and $\mathbf{c}_2$ can be defined as:

\begin{equation}\label{eq:TFHE_framework}
\mathbf{c}_r = F(L(\mathbf{c}_1,\mathbf{c}_2)),
\end{equation}
where $L(.)$ represents the linear operations, e.g. addition and scalar multiplication, and $F(.)$ represents the non-linear function evaluated using a Look-up table (LUT) in the bootstrapping process. As indicated by (\ref{eq:enc_1}), linear operations between ciphertexts accumulate more noise in the resulting ciphertext, and the decryption process fails if the noise crosses a defined threshold. The bootstrapping process resolves this issue by refreshing the noise content in the ciphertext, represented as:

\begin{equation}\label{eq:TFHE_std_binary}
    \mathbf{c}_r = \mathtt{R}_{\mathtt{bsk,LUT}}\left(L(\mathbf{c_1}, \mathbf{c}_2)\right),
\end{equation}
where $\mathbf{c}_r$ is the result ciphertext, $\mathtt{bsk}$ is the bootstrapping key, and $\mathtt{LUT}$ represents the non-linear function $F(.)$. 
$\mathbf{c}_r$ experiences two interesting features: its noise content is refreshed to a pre-calculated level that is independent of the noise 
in $\mathbf{c}_1$ and $\mathbf{c}_2$, and has the function represented by $\mathtt{LUT}$ evaluated on-the-fly on the result of $L(\mathbf{c_1}, \mathbf{c}_2)$. 
These features are important for privacy-preserving inference, as nonlinear operations can be evaluated during bootstrapping rather than approximated by low-degree polynomials \cite{chillotti2021programmable}.

\subsection{Deep Differentiable Logic Gate Networks}
\label{Sec: Deep Differentiable Logic-Gate Networks}
DDLGNs replace traditional weighted arithmetic operators with two-input binary logic gates that compute and forward single-bit activations~\cite{petersen2022deep}. As a result, the final network is intrinsically sparse, weightless, and defined entirely by learned logic functions. This fundamentally distinguishes DDLGNs from BNNs, which merely use fixed logical operations (such as XNOR) to approximate real-valued arithmetic rather than learning the logic gates directly~\cite{huang2025privacy}.

Formally, let \(x=(x_1,\dots,x_d)\in\{0,1\}^d\) denote the binary input to a layer. Each neuron \(j\) selects two inputs indexed by \(u_j,v_j\in\{1,\dots,d\}\) and applies a Boolean function \(f_j\), such that
\begin{equation}
f_j:\{0,1\}^2\rightarrow\{0,1\},
\qquad
z_j=f_j(x_{u_j},x_{v_j}).
\end{equation}
Thus, \((u_j,v_j)\) determines which activations are selected, while \(f_j\) determines how they are combined.
Since a two-input Boolean function is fully specified by four truth-table entries, the set of admissible functions satisfies
\begin{equation}
|\mathcal{F}|=2^{2^2}=16,
\label{eq:logic-16}
\end{equation}
where \(\mathcal{F}\) contains the two constants, the two projections and their negations, and the 10 remaining binary functions, including AND, OR, XOR, XNOR, NAND, NOR, and implication-type gates.

The discrete Boolean operators are not differentiable. Petersen \emph{et al.} \cite{petersen2022deep} address this by using continuous activations in \([0,1]\) during training and allowing each neuron to learn a distribution over the 16 candidate functions. Let \(f^{(i)}\in\mathcal{F}\), \(i\in\{0,\dots,15\}\), denote the \(i\)-th candidate function, and let \(\mathbf{w}^{(j)}\in\mathbb{R}^{16}\) be the trainable score vector of neuron \(j\). The corresponding probabilities and relaxed output are
\begin{equation}
p^{(j)}_i=
\frac{\exp(w^{(j)}_i)}
{\sum_{\ell=0}^{15}\exp(w^{(j)}_\ell)},
\qquad
z_j=
\sum_{i=0}^{15}p^{(j)}_i
f^{(i)}(x_{u_j},x_{v_j}).
\end{equation}
This relaxation enables gradient-based training. After training, each neuron is discretized by selecting the most likely function,
\begin{equation}
f_j^\star=\arg\max_{f^{(i)}\in\mathcal{F}}p^{(j)}_i.
\end{equation}

\begin{figure}[htbp]
    \centering
    \includegraphics[width=0.5\linewidth]{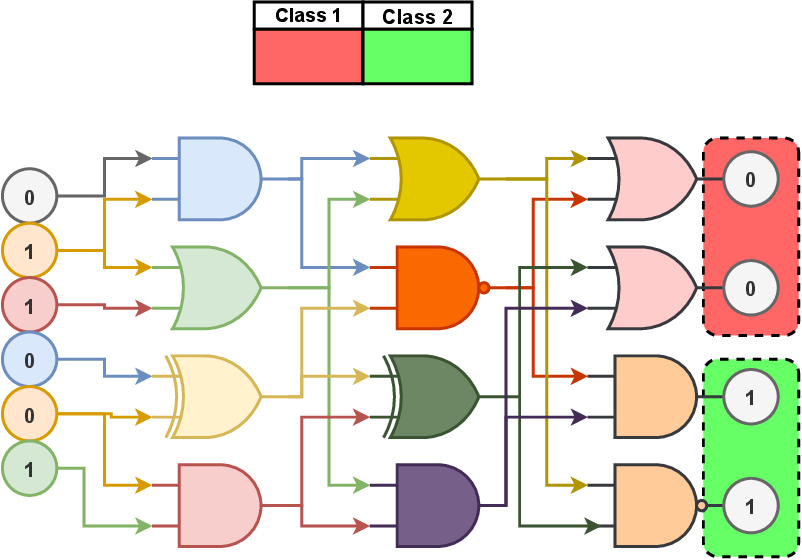}
    \caption{Logic gate network inference with binary activations propagated through learned Boolean gates.}
    \label{fig:LGN}
\end{figure}

For classification, logic gate networks assign multiple output neurons to each class to obtain graded class scores rather than a single binary decision. The \(n_{\mathrm{out}}\) output neurons are divided evenly among \(k\) classes, and the active output bits within each group are aggregated and normalized by a temperature parameter \(\tau\)~\cite{petersen2022deep}. After discretization, this leads to counting the active bits assigned to each class, and the predicted label is obtained from the largest class score. At this stage, inference reduces to propagating binary activations through successive two-input Boolean gates. Fig.~\ref{fig:LGN} illustrates this process using an arbitrary logic gate network.
\FloatBarrier

Compared with arithmetic NNs, or even discretized weight-based models, this produces a representation whose computation is much more directly aligned with TFHE's Boolean execution model. Encrypted inference can therefore be viewed as the homomorphic evaluation of a learned Boolean network.

This Boolean-native formulation has important implications for encrypted inference. Unlike arithmetic neural networks and BNNs, whose encrypted execution remains dominated by weighted aggregations and accumulation structures, discretized DDLGNs reduce inference to the evaluation of learned Boolean functions. Consequently, encrypted execution can be analyzed directly in terms of Boolean-function distributions and gate-level TFHE operations. This observation motivates the PBS cost characterization and execution analysis developed in the following sections.

\section{EI-DDLGN Framework and Threat Model}
\label{Sec:Proposed Framework and Threat Model}

This section describes the deployment workflow of EI-DDLGN and the threat model considered throughout the paper. We consider a client--server inference scenario in which the client wishes to obtain predictions from a trained DDLGN model while keeping the input data private.

\subsection{EI-DDLGN Framework Overview}

\begin{figure}[!ht]
    \centering
    \includegraphics[width=\linewidth]{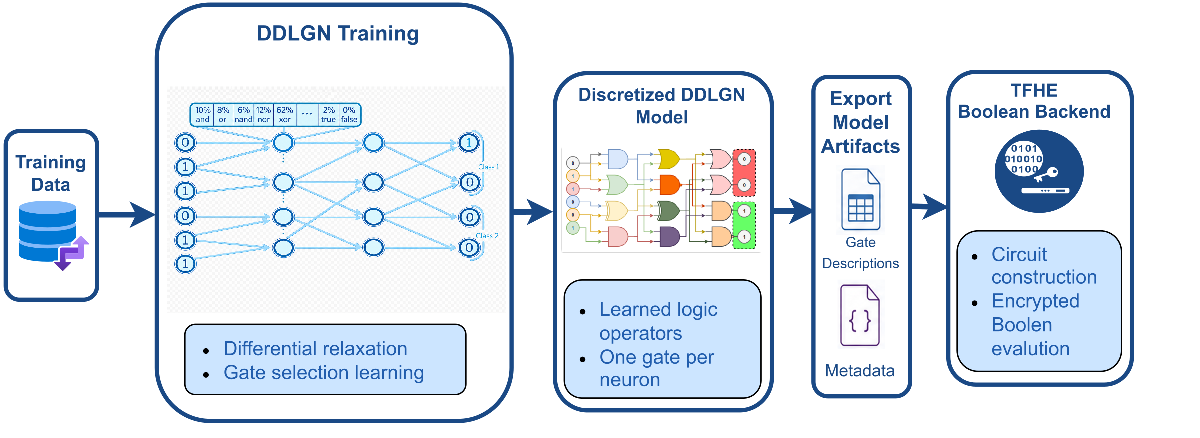}
    \caption{Server-side EI-DDLGN model preparation: training, discretization, export, and TFHE Boolean backend construction.}
    \label{fig:model_preparation}
\end{figure}

\begin{figure}[!ht]
    \centering
    \includegraphics[width=0.48\linewidth]{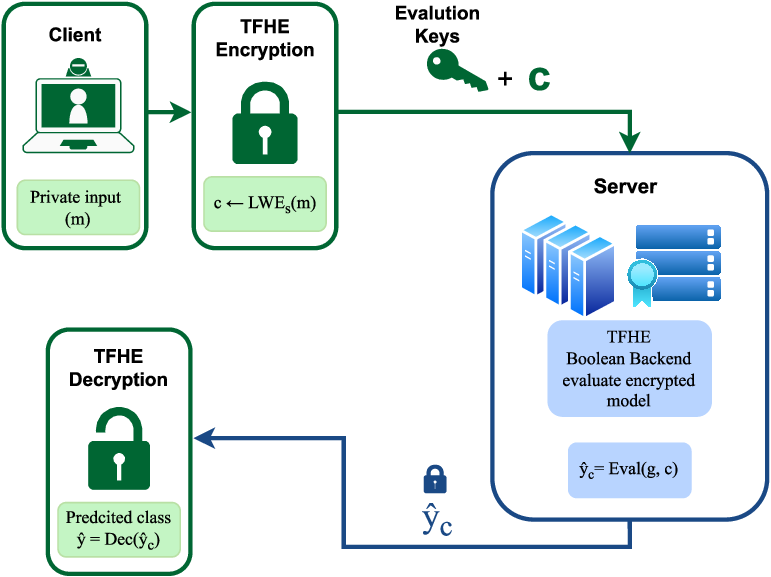}
    \caption{Client--server EI-DDLGN inference under TFHE: the client encrypts inputs and decrypts predictions, while the server evaluates encrypted logic gates.}
    \label{fig:threat_model}
\end{figure}

As illustrated in Figs.~\ref{fig:model_preparation} and~\ref{fig:threat_model}, EI-DDLGN follows a standard encrypted inference workflow consisting of model preparation, client-side encryption, server-side evaluation, and client-side decryption.

The server first trains and discretizes a DDLGN according to the procedure described in Section~\ref{Sec: Deep Differentiable Logic-Gate Networks}. The resulting model is exported as a logic gate network and instantiated using the TFHE Boolean backend. The client generates the TFHE secret and evaluation keys, encrypts the input sample \(m\) into a ciphertext vector \(\mathbf{c}\), retains the secret key locally, and sends the encrypted input together with the evaluation key to the server.

Using the exported logic gate network and the encrypted input, the server evaluates the model directly in the encrypted domain:

\begin{equation}
\hat{\mathbf{y}}_c
=
\mathsf{Eval}(g,\mathbf{c}),
\end{equation}

where \(g\) denotes the deployed DDLGN and \(\hat{\mathbf{y}}_c\) the encrypted prediction output. Evaluation proceeds layer-by-layer by propagating ciphertext activations through the learned Boolean gates. Since all gates within a layer depend only on activations from the previous layer, they can be evaluated independently and executed in parallel. The encrypted prediction is then returned to the client.

Finally, the client decrypts the received ciphertext and obtains

\begin{equation}
\hat{\mathbf{y}} = g(m).
\end{equation}

Correctness requires encrypted execution to produce the same output as plaintext inference:

\begin{equation}
\mathsf{Dec}\!\left(\mathsf{Eval}(g,\mathbf{c})\right)
=
g(m).
\end{equation}

\subsection{Protected Assets and Adversary Model}

The primary objective of EI-DDLGN is to protect the confidentiality of client inputs during inference. Prediction outputs are also protected from the server and remain encrypted throughout server-side evaluation. Only the client can decrypt the final prediction.

We assume an \emph{honest-but-curious} adversary model~\cite{paverd2014modelling}. The server is assumed to execute the protocol correctly but may attempt to infer information about client inputs, intermediate activations, or prediction outputs from the ciphertexts it processes. Throughout inference, all client-dependent inputs and intermediate values remain encrypted under TFHE, and the client retains the secret decryption key locally.

EI-DDLGN does not introduce additional cryptographic assumptions beyond those of TFHE and therefore inherits its security guarantees based on the hardness of the Learning With Errors (LWE) problem. Possession of the evaluation and bootstrapping keys enables homomorphic computation but does not enable recovery of plaintext inputs or outputs.

We do not consider model confidentiality. The server is assumed to know the deployed DDLGN architecture, the learned Boolean functions, and all associated model metadata. We further exclude malicious protocol deviations, denial-of-service attacks, incorrect computation, and side-channel attacks based on timing, power consumption, cache behavior, or hardware leakage.

Under these assumptions, privacy-preserving inference reduces to securely evaluating \(\mathsf{Eval}(g,\mathbf{c})\) such that the server learns no information about the client's plaintext input while the client alone can recover the final prediction.

\section{Cost Structure of TFHE-Based DDLGN Inference}
\label{sec:pbs-analysis}

This section analyzes the cost structure of TFHE-based DDLGN inference.
Unlike arithmetic neural networks, whose encrypted execution is primarily
governed by accumulator growth and circuit bit-width, DDLGNs execute as
learned Boolean networks whose cost depends on the selected Boolean
functions and their interaction with TFHE operations.

We first characterize the PBS requirements of the learned Boolean
functions. We then introduce Model-Fixed-Wire PBS Bypass
(MFW-PBS Bypass), a semantics-preserving execution strategy that avoids
unnecessary PBS operations. Finally, we analyze how learned gate
distributions, propagated wire status, and network topology affect the
executed PBS budget and the resulting accuracy--cost trade-offs.

\subsection{PBS Cost Characterization}
\label{sec:pbs-cost-characterization}

Based on Eq.~(\ref{eq:logic-16}), each EI-DDLGN neuron implements one of
the 16 two-input Boolean functions. However, these functions do not have
the same encrypted cost under TFHE. We classify the functions requiring
a binary PBS as

\begin{equation}
\mathcal{G}_{\mathrm{PBS}}
=
\{
\mathsf{AND},\mathsf{OR},\mathsf{XOR},\mathsf{XNOR},
\mathsf{NAND},\mathsf{NOR},
\mathsf{IMP},\mathsf{RIMP},\mathsf{NIMP},\mathsf{NRIMP}
\},
\label{eq:pbs-gates}
\end{equation}

where \(\mathsf{IMP}\), \(\mathsf{RIMP}\), \(\mathsf{NIMP}\), and
\(\mathsf{NRIMP}\) denote implication, reverse implication, and their
negations~\cite{TFHE-rs}. The remaining six functions,

\begin{equation}
\mathcal{G}_{\mathrm{noPBS}}
=
\{\mathbf{0},\mathbf{1},a,b,\neg a,\neg b\},
\label{eq:nopbs-gates}
\end{equation}

can be implemented as model-fixed outputs, ciphertext forwarding, or
encrypted negation without invoking a binary PBS.

Accordingly, the baseline PBS cost of a learned function \(f\) can be
written as

\begin{equation}
\operatorname{Cost}_{\mathrm{PBS}}(f)
=
\mathds{1}\{f\in\mathcal{G}_{\mathrm{PBS}}\}.
\label{eq:gate-pbs-cost}
\end{equation}

Thus, two EI-DDLGNs with the same number of neurons may have different
encrypted costs depending on the Boolean functions selected during
training.

Gate type alone, however, does not fully determine the final PBS cost.
During layer-wise inference, some intermediate wires may become
input-independent and therefore constant for every possible client
input. Substituting such values into functions in
\(\mathcal{G}_{\mathrm{PBS}}\) can reduce them to constants,
projections, or negated projections, eliminating the need for a binary
PBS. This observation motivates the execution strategy presented next.

\subsection{Model-Fixed-Wire PBS Bypass}
\label{sec:mfw-pbs-bypass}

A model-fixed wire is a wire whose value is constant for all possible
client inputs. Since its value is known independently of the encrypted
input, it can be substituted directly into the truth table of a learned
Boolean function. In many cases, this reduces a PBS-requiring gate to a
simpler operation such as a constant output, ciphertext forwarding, or
encrypted negation. MFW-PBS Bypass exploits this property by selecting
the simplest equivalent TFHE operation for each gate during encrypted
evaluation.

During model preparation, the truth table of every learned gate is used to determine the TFHE action to be executed for each possible
model-fixed input state. During encrypted inference, the server
evaluates the DDLGN layer-by-layer and executes the precomputed action
dictated by the propagated wire-status metadata. The network topology
remains unchanged; the bypass modifies only the execution operator used
for a gate and never inspects, decrypts, or branches on plaintext data.

Each wire \(w\) is assigned a status

\begin{equation}
\sigma(w)
\in
\{\mathsf{M}_0,\mathsf{M}_1,\mathsf{E}\},
\label{eq:wire-status}
\end{equation}

where \(\mathsf{M}_0\) and \(\mathsf{M}_1\) denote model-fixed
input-independent bits and \(\mathsf{E}\) denotes a client-dependent
encrypted bit. All original input wires are initialized as

\begin{equation}
\sigma(x_i)=\mathsf{E}.
\label{eq:input-status}
\end{equation}

A wire is assigned \(\mathsf{M}_{b_{\mathrm{fix}}}\) only when its
value is identical for all possible client inputs:

\begin{equation}
\sigma(w)=\mathsf{M}_{b_{\mathrm{fix}}}
\quad\Longrightarrow\quad
\forall\,m\in\mathcal{X},
\;
w(m)=b_{\mathrm{fix}},
\qquad
b_{\mathrm{fix}}\in\{0,1\},
\label{eq:model-fixed-invariant}
\end{equation}

where \(\mathcal{X}\) denotes the client-input domain. Therefore, no
client-dependent input, intermediate value, or prediction component is
classified as model-fixed.

For a gate \(f(a,b)\), the value of each model-fixed input is
substituted into the gate's truth table. The resulting effective Boolean
function \(\widetilde{f}\) satisfies

\begin{equation}
\widetilde{f}
\in
\{\mathbf{0},\mathbf{1},x,\neg x\}
\cup
\mathcal{G}_{\mathrm{PBS}}.
\label{eq:residual-functions}
\end{equation}

The server then selects the corresponding operation:

\begin{equation}
\mathsf{Act}(\widetilde{f})=
\begin{cases}
\mathsf{M}_{b_{\mathrm{fix}}},
&
\widetilde{f}=b_{\mathrm{fix}},
\\[2mm]
\mathbf{c}_x,
&
\widetilde{f}=x,
\\[2mm]
\mathsf{NOT}(\mathbf{c}_x),
&
\widetilde{f}=\neg x,
\\[2mm]
\mathsf{PBS}_{\widetilde{f}}(\mathbf{c}_a,\mathbf{c}_b),
&
\widetilde{f}\in\mathcal{G}_{\mathrm{PBS}}.
\end{cases}
\label{eq:mfw-pbs-action}
\end{equation}

Thus, each neuron is evaluated using one of four actions: a model-fixed
output, ciphertext forwarding, encrypted negation, or a binary PBS.
Only the final case requires an actual PBS operation.

Table~\ref{tab:model-fixed-collapse} summarizes the resulting actions
for the ten PBS-requiring Boolean functions when one input is
model-fixed.

\begin{table}[H]
\centering
\caption{TFHE action selected when one input to a PBS-requiring Boolean function is model-fixed. Here, \(x\) denotes the remaining encrypted input, while \(0\) and \(1\) denote model-fixed outputs.}
\label{tab:model-fixed-collapse}
\setlength{\tabcolsep}{5pt}
\begin{tabular}{lcccc}
\toprule
Gate \(f(a,b)\) & \(f(0,x)\) & \(f(1,x)\) & \(f(x,0)\) & \(f(x,1)\) \\
\midrule
\(a\land b\)              & 0          & \(x\)       & 0          & \(x\) \\
\(a\lor b\)               & \(x\)      & 1           & \(x\)      & 1 \\
\(a\oplus b\)             & \(x\)      & \(\neg x\)  & \(x\)      & \(\neg x\) \\
\(\neg(a\oplus b)\)       & \(\neg x\) & \(x\)       & \(\neg x\) & \(x\) \\
\(\neg(a\land b)\)        & 1          & \(\neg x\)  & 1          & \(\neg x\) \\
\(\neg(a\lor b)\)         & \(\neg x\) & 0           & \(\neg x\) & 0 \\
\(a\Rightarrow b\)        & 1          & \(x\)       & \(\neg x\) & 1 \\
\(a\Leftarrow b\)         & \(\neg x\) & 1           & 1          & \(x\) \\
\(\neg(a\Rightarrow b)\)  & 0          & \(\neg x\)  & \(x\)      & 0 \\
\(\neg(a\Leftarrow b)\)   & \(x\)      & 0           & 0          & \(\neg x\) \\
\bottomrule
\end{tabular}
\end{table}

Because model-fixed inputs are substituted directly into the truth table
of each learned Boolean function, the selected TFHE action implements an
equivalent Boolean mapping and therefore preserves inference
correctness.

The property in Eq.~(\ref{eq:model-fixed-invariant}) is preserved
throughout layer-wise inference. A model-fixed output is produced only
when the effective Boolean function is constant for every client input.
Ciphertext forwarding and encrypted negation preserve the encrypted
representation of client-dependent values, while the remaining binary
gates operate directly on ciphertexts. Consequently, all
client-dependent intermediate values and prediction scores remain
encrypted throughout server-side inference.

\subsection{PBS Cost Model and Metrics}
\label{sec:pbs-cost-model}

We next define the metrics used to quantify PBS cost before and after
applying MFW-PBS Bypass. Let
\(\mathcal{G}_g\) denote the set of gates in model \(g\), and let
\(f_{\gamma}\) denote the Boolean function implemented by gate
\(\gamma\in\mathcal{G}_g\). The total number of gates is
\begin{equation}
G(g)=|\mathcal{G}_g|.
\label{eq:total-gates}
\end{equation}

The baseline assumes that all gate inputs are encrypted and counts every
learned function that requires a binary PBS:
\begin{equation}
N_{\mathrm{PBS}}^{\mathrm{base}}(g)
=
\sum_{\gamma\in\mathcal{G}_g}
\mathds{1}
\left\{
f_{\gamma}\in\mathcal{G}_{\mathrm{PBS}}
\right\}.
\label{eq:npbs-base}
\end{equation}

Under MFW-PBS Bypass, a PBS is executed only
when the learned function belongs to \(\mathcal{G}_{\mathrm{PBS}}\) and both
of its input wires are not model-fixed bits. The executed PBS count is therefore
\begin{equation}
N_{\mathrm{PBS}}^{\mathrm{exec}}(g)
=
\sum_{\gamma\in\mathcal{G}_g}
\mathds{1}
\left\{
f_{\gamma}\in\mathcal{G}_{\mathrm{PBS}},
\;
\sigma(a_{\gamma})=\mathsf{E},
\;
\sigma(b_{\gamma})=\mathsf{E}
\right\},
\label{eq:npbs-exec}
\end{equation}
where \(a_{\gamma}\) and \(b_{\gamma}\) denote the two input wires of gate
\(\gamma\).

The PBS bypass rate achieved for model \(g\) is
\begin{equation}
\beta_{\mathrm{PBS}}(g)
=
100\cdot
\frac{
N_{\mathrm{PBS}}^{\mathrm{base}}(g)
-
N_{\mathrm{PBS}}^{\mathrm{exec}}(g)
}{
N_{\mathrm{PBS}}^{\mathrm{base}}(g)
}.
\label{eq:pbs-bypass-rate}
\end{equation}

To express the PBS cost relative to the complete network size, we further
define the baseline and executed PBS shares as:
\begin{equation}
\rho_{\mathrm{PBS}}^{\mathrm{base}}(g)
=
100\cdot
\frac{N_{\mathrm{PBS}}^{\mathrm{base}}(g)}{G(g)},
\qquad
\rho_{\mathrm{PBS}}^{\mathrm{exec}}(g)
=
100\cdot
\frac{N_{\mathrm{PBS}}^{\mathrm{exec}}(g)}{G(g)}.
\label{eq:pbs-shares}
\end{equation}

Finally, let \(y_{\gamma}\) denote the output wire of gate \(\gamma\). The
percentage of gates resolved to model-fixed outputs is
\begin{equation}
\rho_{\mathrm{fixed}}(g)
=
100\cdot
\frac{
\displaystyle
\sum_{\gamma\in\mathcal{G}_g}
\mathds{1}
\left\{
\sigma(y_{\gamma})
\in
\{\mathsf{M}_0,\mathsf{M}_1\}
\right\}
}{
G(g)
}.
\label{eq:fixed-share}
\end{equation}

Generally,
\(N_{\mathrm{PBS}}^{\mathrm{exec}}\)
gives the remaining number of PBS operations, while
\(\rho_{\mathrm{PBS}}^{\mathrm{exec}}\)
measures their share of the complete network. In contrast,
\(\beta_{\mathrm{PBS}}\)
measures the fraction of baseline PBS operations bypassed, and
\(\rho_{\mathrm{fixed}}\)
measures the extent of model-fixed propagation.

\subsection{PBS Bypass Across Model Architectures}
\label{sec:pbs-bypass-analysis}

We analyze the behavior of MFW-PBS Bypass across the complete EI-DDLGN
model grid. Let \(\mathcal{M}\) denote the set of 72 models constructed
from three datasets, six depths, and four widths. The datasets are
MNIST~\cite{deng2012mnist}, FashionMNIST~\cite{xiao2017fashion}, and UCI
Phishing Websites~\cite{mohammad2015phishing}, with
\begin{equation}
d\in\{1,2,3,4,5,6\},
\qquad
w\in\{2000,4000,6000,8000\}.
\label{eq:pbs-analysis-grid}
\end{equation}
Thus, each dataset contributes 24 depth and width configurations. This grid enables us to examine how model size, learned gate
composition, and model-fixed wire propagation affect the executed PBS budget.

\begin{figure}[!htbp]
    \centering
    \includegraphics[width=\linewidth]
    {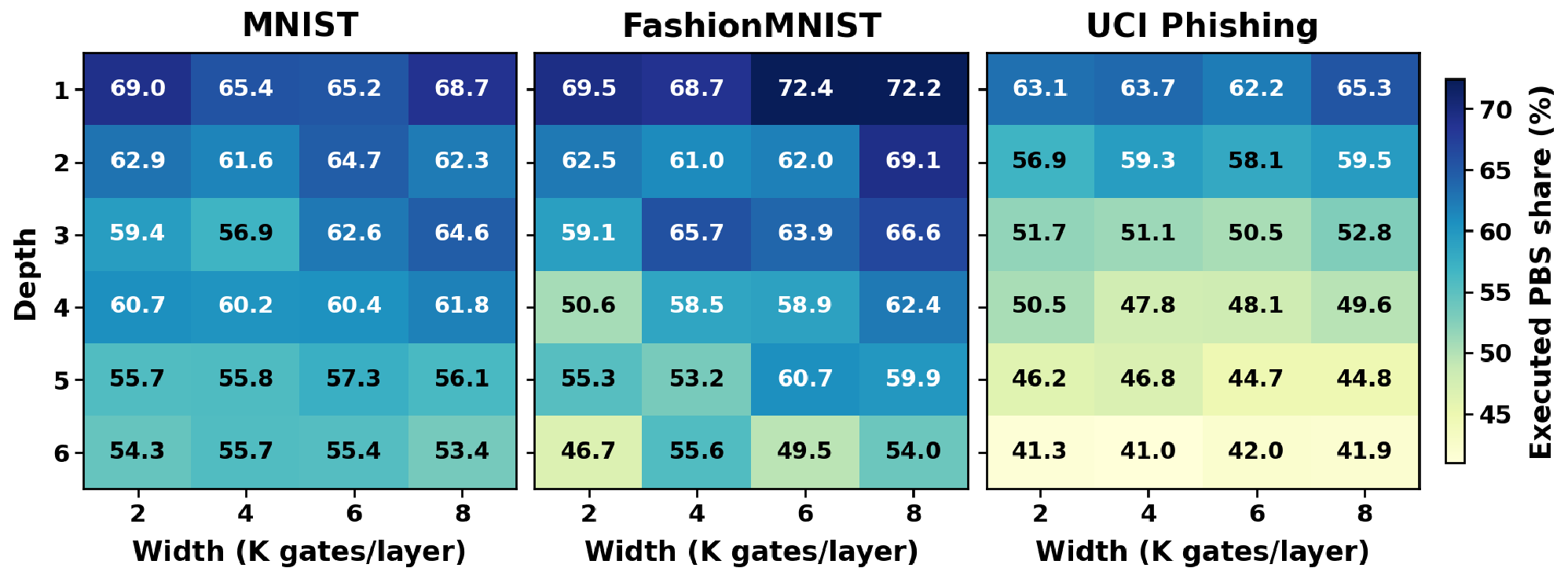}
    \caption{\(\rho_{\mathrm{PBS}}^{\mathrm{exec}}\) across the EI-DDLGN
    depth--width grid under MFW-PBS Bypass. Each panel corresponds to one dataset.}
    \label{fig:fps_executed_pbs_share_heatmap}
\end{figure}

As shown in Table~\ref{tab:fps-selected-ei-ddlgn-pbs},
MFW-PBS Bypass reduces the number of executed PBS operations for every
selected configuration. However, the percentage of model-fixed outputs
and the resulting bypass rate do not necessarily change proportionally.
The final reduction depends not only on how many model-fixed wires are
produced, but also on where they occur and which downstream Boolean
functions receive them. Figure~\ref{fig:fps_executed_pbs_share_heatmap} shows the executed PBS share after applying MFW-PBS Bypass.

\begin{table}[htpb!]
\centering
\caption{PBS characteristics of representative EI-DDLGN models.}
\label{tab:fps-selected-ei-ddlgn-pbs}
\setlength{\tabcolsep}{5pt}
\resizebox{\linewidth}{!}{%
\begin{tabular}{llrrrrr}
\toprule
Model
& Dataset
& Acc. (\%)
& \(\rho_{\mathrm{fixed}}\) (\%)
& \(\rho_{\mathrm{PBS}}^{\mathrm{base}}\) (\%)
& \(\rho_{\mathrm{PBS}}^{\mathrm{exec}}\) (\%)
& \(\beta_{\mathrm{PBS}}\) (\%) \\
\midrule

\multirow{3}{*}{\makecell[l]{EI-DDLGN (Small)\\
\((2\times4K)\)\\
\(\tau=3\)}}
& MNIST        & 91.55 & 4.59 & 65.22 & 61.58 & 5.60 \\
& FashionMNIST & 77.49 & 4.99 & 63.96 & 61.00 & 4.63 \\
& UCI Phishing & 92.40 & 9.85 & 64.45 & 59.30 & 7.99 \\
\cmidrule(lr){1-7}

\multirow{3}{*}{\makecell[l]{EI-DDLGN (Medium)\\
\((4\times6K)\)\\
\(\tau=5\)}}
& MNIST        & 95.87 & 2.69 & 63.58 & 60.44 & 4.94 \\
& FashionMNIST & 81.27 & 2.15 & 61.03 & 58.91 & 3.48 \\
& UCI Phishing & 93.93 & 9.38 & 58.83 & 48.08 & 18.27 \\
\cmidrule(lr){1-7}

\multirow{3}{*}{\makecell[l]{EI-DDLGN (Large)\\
\((6\times8K)\)\\
\(\tau=10\)}}
& MNIST        & 97.20 & 4.34 & 59.22 & 53.36 & 9.91 \\
& FashionMNIST & 83.51 & 3.78 & 58.47 & 53.99 & 7.66 \\
& UCI Phishing & 95.23 & 9.10 & 55.55 & 41.87 & 24.62 \\
\bottomrule
\end{tabular}
}
\end{table}

\begin{figure}[!htbp]
    \centering
    \includegraphics[width=0.7\linewidth]
    {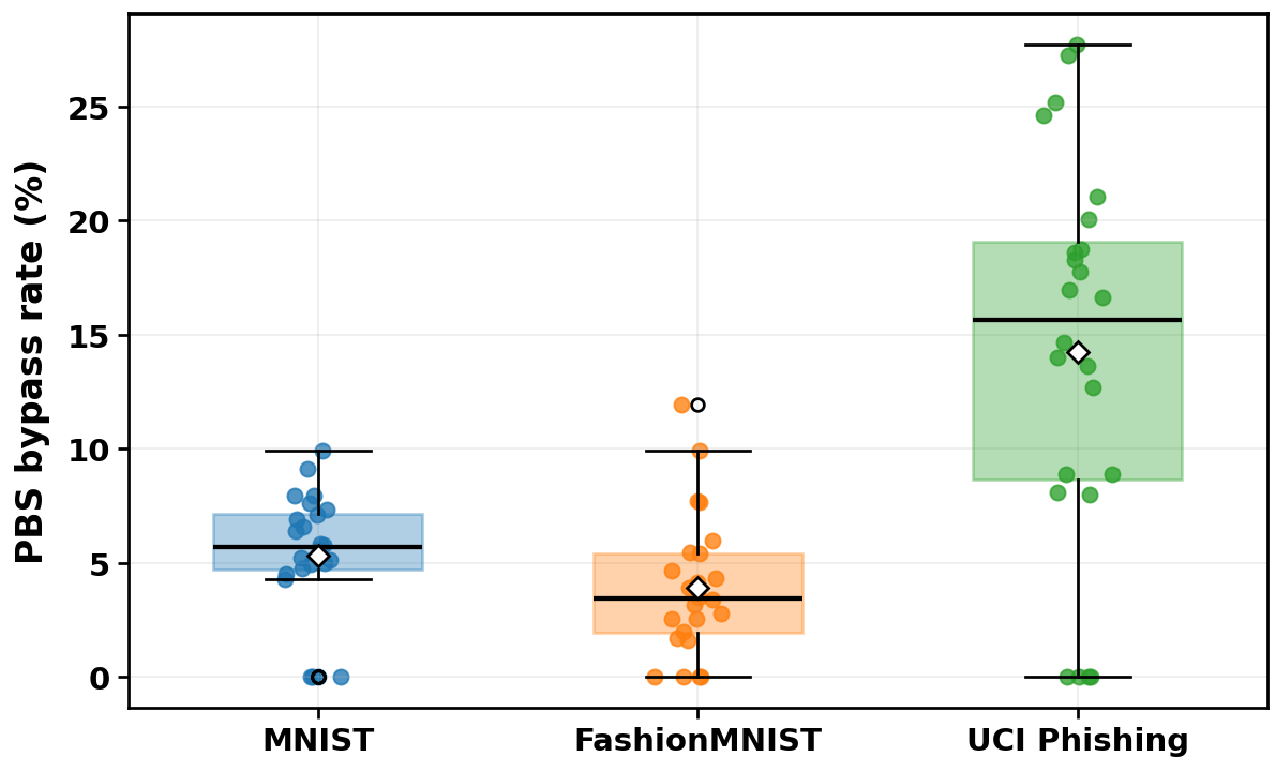}
    \caption{Distribution of the PBS bypass rate
    \(\beta_{\mathrm{PBS}}\) across the analyzed EI-DDLGN models.}
    \label{fig:fps_pbs_bypass_distribution}
\end{figure}

The remaining PBS share depends not only on network size, but also on
the learned Boolean-function distribution and the propagation of
model-fixed wires across layers. Consequently, models with the same
depth and width may exhibit different PBS shares when trained on
different datasets. Figure~\ref{fig:fps_pbs_bypass_distribution} shows the distribution of
\(\beta_{\mathrm{PBS}}\) over the complete model grid.

The mean PBS bypass rates are \(5.32\%\) for MNIST,
\(3.92\%\) for FashionMNIST, and \(14.23\%\) for UCI Phishing. The
corresponding maximum rates are \(9.91\%\), \(11.91\%\), and
\(27.69\%\), respectively.

The zero-bypass cases correspond to depth-1 models. Since gates within
the same layer are evaluated independently, no earlier layer exists from
which a model-fixed wire can propagate into a subsequent gate. Deeper
networks therefore provide more opportunities for model-fixed values to
simplify downstream computations.

Overall, the results confirm that the PBS cost of EI-DDLGN is determined
jointly by learned gate distributions, depth, and propagated wire
status, rather than by total gate count alone.

\subsection{Accuracy--PBS Budget Trade-off}
\label{sec:accuracy-pbs-tradeoff}

\begin{figure}[!htbp]
    \centering
    \includegraphics[width=0.7\linewidth]
    {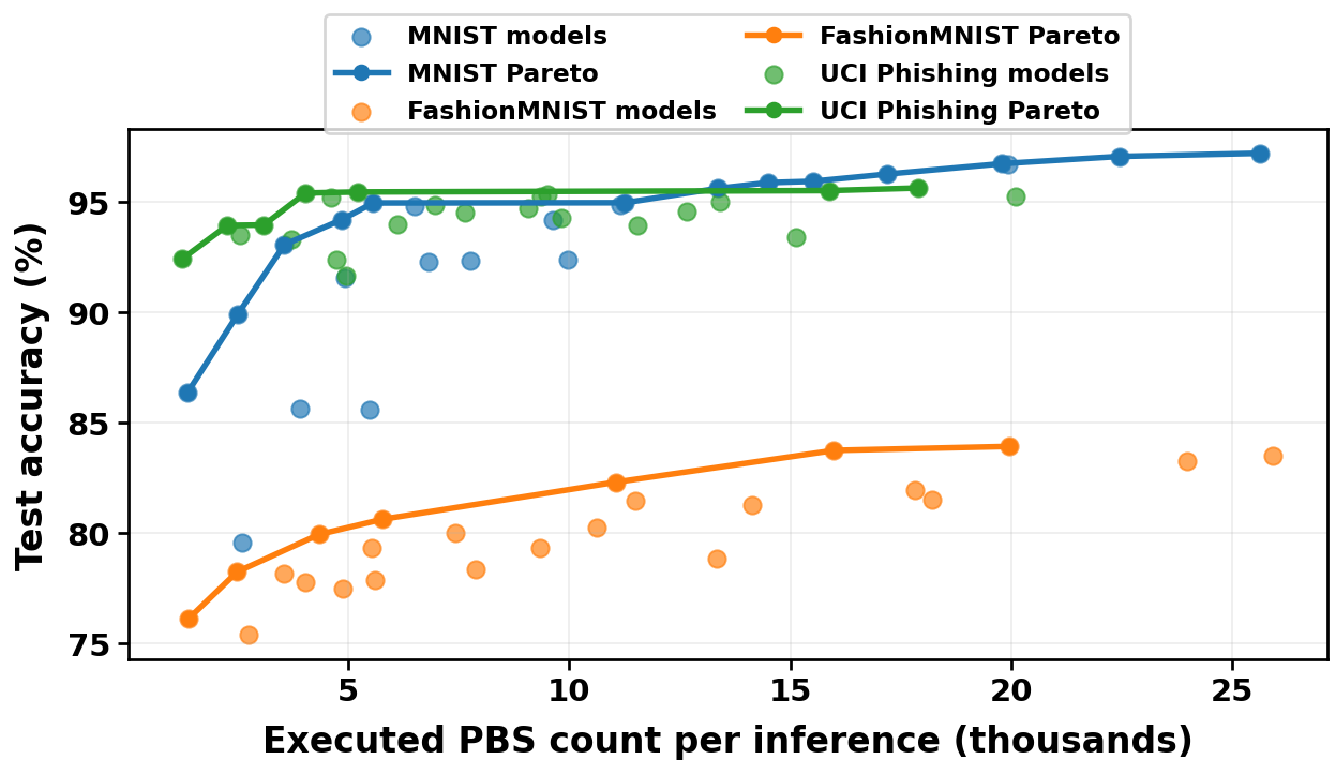}
    \caption{Accuracy versus executed PBS count for the EI-DDLGN model grid.}
    \label{fig:fps_accuracy_vs_executed_pbs}
\end{figure}

We next study the trade-off between prediction accuracy and executed PBS
cost over the complete EI-DDLGN model grid. Figure~\ref{fig:fps_accuracy_vs_executed_pbs}
plots each model using
\begin{equation}
\text{\(x\)-axis: }\frac{N_{\mathrm{PBS}}^{\mathrm{exec}}(g)}{1000},
\qquad
\text{\(y\)-axis: }\mathrm{Acc}(g),
\qquad
g\in\mathcal{M}.
\label{eq:accuracy-pbs-axes}
\end{equation}
The highlighted Pareto frontier contains models for which no other model
on the same dataset achieves both lower executed PBS count and higher or
equal accuracy.

Figure~\ref{fig:fps_accuracy_vs_executed_pbs} shows that increasing the
executed PBS budget generally improves accuracy, although the benefit is
dataset dependent. MNIST continues to benefit from additional PBS budget
over most of the evaluated range, FashionMNIST improves more gradually,
and UCI Phishing reaches high accuracy at relatively modest executed PBS
counts. In all three datasets, accuracy gains eventually diminish as the
executed PBS budget increases.

The same trade-off can be viewed from a budget-constrained perspective.
For a PBS budget \(\Lambda\), the achievable accuracy is defined as
\begin{equation}
B_{\mathrm{PBS}}(\Lambda)
=
\max_{\substack{
g\in\mathcal{M}\\
N_{\mathrm{PBS}}^{\mathrm{exec}}(g)\leq\Lambda
}}
\mathrm{Acc}(g).
\label{eq:pbs-budget}
\end{equation}

\begin{figure}[!htbp]
    \centering
    \includegraphics[width=0.7\linewidth]
    {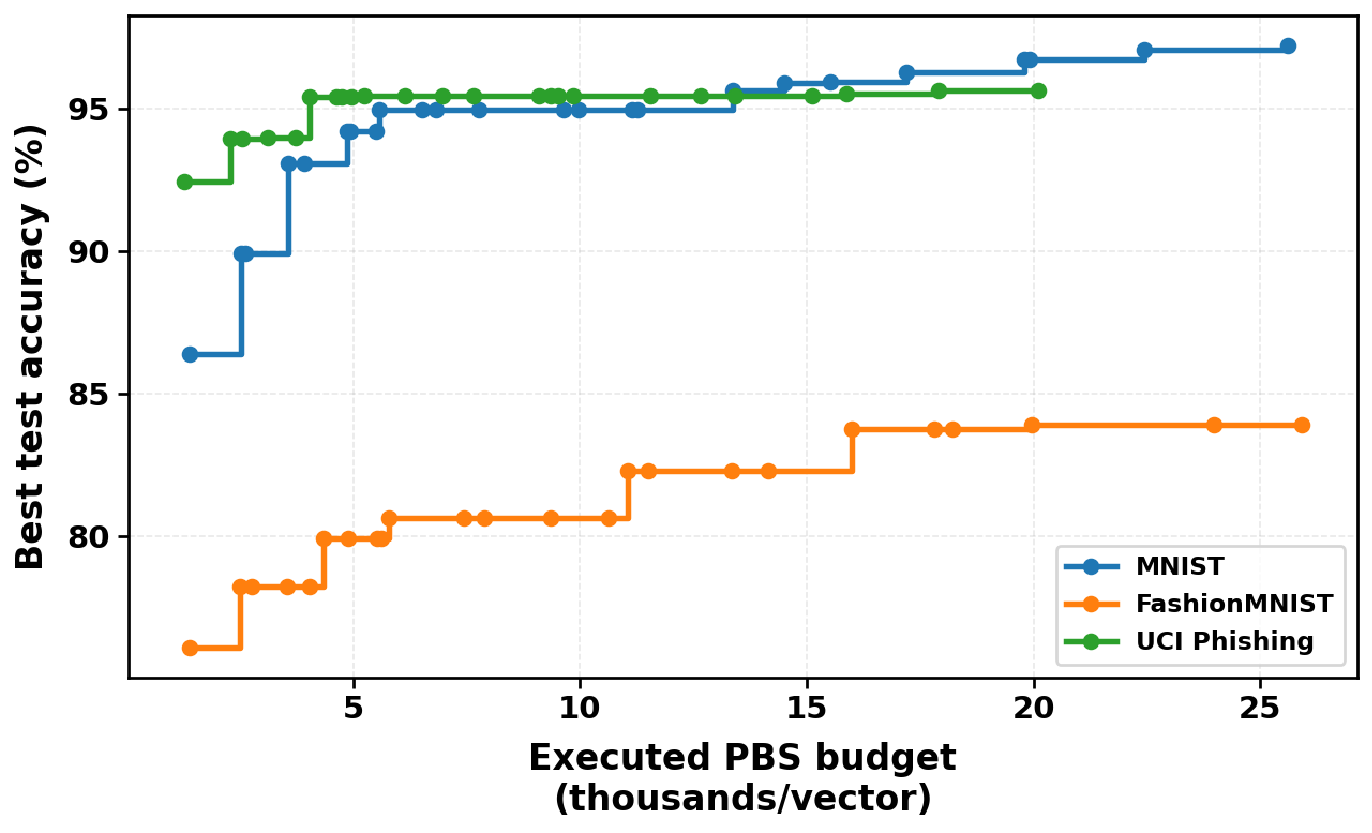}
    \caption{Best accuracy achievable under an EI-DDLGN executed PBS budget.}
    \label{fig:fps_best_accuracy_under_executed_pbs_budget}
\end{figure}

Figure~\ref{fig:fps_best_accuracy_under_executed_pbs_budget} shows the
best accuracy obtainable under a given executed PBS budget. The step-like
behavior arises because the optimal model changes only when a different
configuration becomes the highest-accuracy model within the budget
constraint. The curves further illustrate the diminishing-return trend
observed in Fig.~\ref{fig:fps_accuracy_vs_executed_pbs}: increasing the
PBS budget is most beneficial at low budgets, while additional encrypted
computation provides progressively smaller accuracy improvements at
higher budgets.

\subsection{Discussion}
The preceding analysis highlights an important difference between
EI-DDLGN and arithmetic TFHE inference. In arithmetic neural networks,
encrypted execution cost is largely determined by accumulator growth,
circuit bit-width, and programmable bootstrapping requirements. In
contrast, EI-DDLGN expresses inference directly as a learned Boolean
computation. Consequently, encrypted cost depends not only on network
size, but also on the learned Boolean-function distribution and the
propagated status of intermediate wires.

These results suggest that model architecture plays a central role in
determining encrypted inference efficiency under TFHE and motivate the
exploration of other Boolean-native learning architectures.

\section{Encrypted Inference Time Analysis}
\label{sec:encrypted-inference-time}

Building on the PBS cost analysis in Section~\ref{sec:pbs-analysis}, we now
analyze how the learned network structure and the remaining PBS operations
affect encrypted inference time. After training, EI-DDLGN inference
is the homomorphic evaluation of a layered logic gate network. Let the discretized DDLGN contain \(L\) logic gate layers, and let \(n_\ell\) denote the number of neurons in layer \(\ell\). Since each neuron corresponds to one two-input Boolean operator, the total number of logic gates is
\begin{equation}
G(g) =\sum_{\ell=1}^{L}n_\ell,
\qquad
G(g)=Ln \quad \text{for a uniform width network.}
\end{equation}

Thus, under fixed TFHE backend parameters, the worst-case encrypted evaluation cost 
grows as \(T_{\mathrm{eval}}=\Theta(G)\). However, the practical cost is governed more directly by 
the number of PBS operations executed after MFW-PBS Bypass, described 
in Section~\ref{sec:mfw-pbs-bypass}. Hence, the executed PBS count \(N_{\mathrm{PBS}}^{\mathrm{exec}}(g)\), rather than only the raw gate count \(G(g)\), determines how many expensive binary TFHE Boolean operations 
remain after propagation.
Moreover, the encrypted evaluation time can be written layer-wise as
\begin{equation}
T_{\mathrm{eval}}(g)
\approx
\sum_{\ell=1}^{L} T_{\mathrm{eval},\ell}(g),
\qquad
T_{\mathrm{eval},\ell}(g) \propto N_{\mathrm{PBS},\ell}^{\mathrm{exec}}(g),
\label{eq:eval-time-layerwise}
\end{equation}
where \(T_{\mathrm{eval},\ell}(g)\) denotes the encrypted evaluation time of layer 
\(\ell\), and \(N_{\mathrm{PBS},\ell}^{\mathrm{exec}}(g)\) is the executed PBS count in that layer.

The layered structure also determines parallelism. Gates inside the same layer can be evaluated independently. 
Therefore, wider networks benefit more from intra-layer parallelization, whereas deeper networks increase the number 
of sequential rounds. 

The end-to-end encrypted inference time is modeled as

\begin{equation}
T_{\mathrm{total}}(g,m)
=
T_{\mathrm{enc}}(m)
+
T_{\mathrm{eval}}(g)
+
T_{\mathrm{dec}}(\mathbf{c}_{\mathrm{out}}),
\label{eq:total-time}
\end{equation}

where \(T_{\mathrm{enc}}(m)\), \(T_{\mathrm{eval}}(g)\), and
\(T_{\mathrm{dec}}(\mathbf{c}_{\mathrm{out}})\) denote the input encryption,
server evaluation, and output decryption times, respectively. In our setting,
\(T_{\mathrm{eval}}(g)\) is expected to dominate because it includes the
remaining PBS operations across all network layers.

The latency model highlights an important distinction between
EI-DDLGN and arithmetic TFHE inference. Since inference is expressed
directly as a layered Boolean computation, hidden-layer execution
does not require arithmetic accumulation. Consequently, encrypted
execution cost is governed more directly by the learned logic
structure and the executed PBS count.

The empirical evaluation in Section~\ref{sec:encrypted-evaluation}
validates this model and examines how these factors translate into
practical encrypted inference latency.

\section{Encrypted Evaluation and Baseline Comparison}
\label{sec:encrypted-evaluation}
This section evaluates the end-to-end encrypted execution of EI-DDLGN and
validates the inference time analysis presented in
Section~\ref{sec:encrypted-inference-time}. We first describe the
implementation, cryptographic parameters, and hardware platform. We then analyze encrypted time scaling across model depth and
width and compare EI-DDLGN with arithmetic-based QAT-FCNN baselines.

\subsection{Experimental Setup}
\label{sec:experimental-setup}

We implemented EI-DDLGN training, discretization, and model export in Python using \texttt{difflogic}~\cite{petersen2022deep}. Models were trained with cross-entropy loss and Adam ($\text{lr}=0.01$, batch size 100 for MNIST/Fashion-MNIST, 256 for UCI Phishing) using early stopping with patience 10 on validation accuracy. Unless otherwise stated, the
same training procedure was used across all evaluated configurations. DDLGN architectures require binary-valued inputs. Therefore, images were thresholded at $0.5$ into 784-bit vectors. The 30 ternary UCI Phishing features were one-hot encoded into 90-bit vectors, with labels mapped to $\{0, 1\}$.

We implemented the loading and encrypted evaluation of
the exported Boolean gate networks using a backend in Rust with the help of \texttt{TFHE} library~\cite{TFHE-rs}. 
Moreover, the cryptographic parameters used in EI-DDLGN are summarized in Table~\ref{tab:tfhe-parameters}. These cryptographic parameters are adapted from the default parameters in~\cite{TFHE-rs}. We also evaluated the security level of this parameter set using the Lattice
Estimator~\cite{albrecht2021lattice}, which reports a minimum estimated
security level of 132 bits against known lattice attacks. 

The encrypted time analysis uses the same model grid
\(\mathcal{M}\) introduced in Section~\ref{sec:pbs-bypass-analysis},
consisting of 72 EI-DDLGN models across the same three datasets. The Small, Medium, and Large configurations used in the
representative comparisons correspond to \(2\times4K\), \(4\times6K\), and
\(6\times8K\), with temperatures \(\tau=3\), \(\tau=5\), and \(\tau=10\),
respectively. These configurations were selected as representative points spanning
the accuracy--latency trade-off observed across the full model grid. All encrypted evaluations, including EI-DDLGN and FCNN baseline measurements,
were conducted on the same desktop platform equipped with an Intel Core
i9-10900 processor with 10 cores and 20 threads at 2.8\,GHz, and 32\,GB of DDR4 RAM.

\begin{table}[ht]
\centering
\caption{TFHE parameters used in the encrypted experiments.}
\label{tab:tfhe-parameters}
\setlength{\tabcolsep}{7pt}
\begin{tabular}{lll}
\toprule
Parameter & Symbol & Value \\
\midrule
LWE dimension
    & \(n\)
    & \(739\) \\
Ciphertext modulus
    & \(q\)
    & \(2^{32}\) \\
LWE noise standard deviation
    & \(\sigma_{\mathrm{LWE}}\)
    & \(5.862\times10^{-6}\) \\
Estimated security level
    & --
    & \(132\) bits \\
\bottomrule
\end{tabular}
\end{table}

\subsection{Encrypted Time Scaling Across Depth and Width}
\begin{figure}[!htbp]
    \centering
    \includegraphics[width=0.9\linewidth]{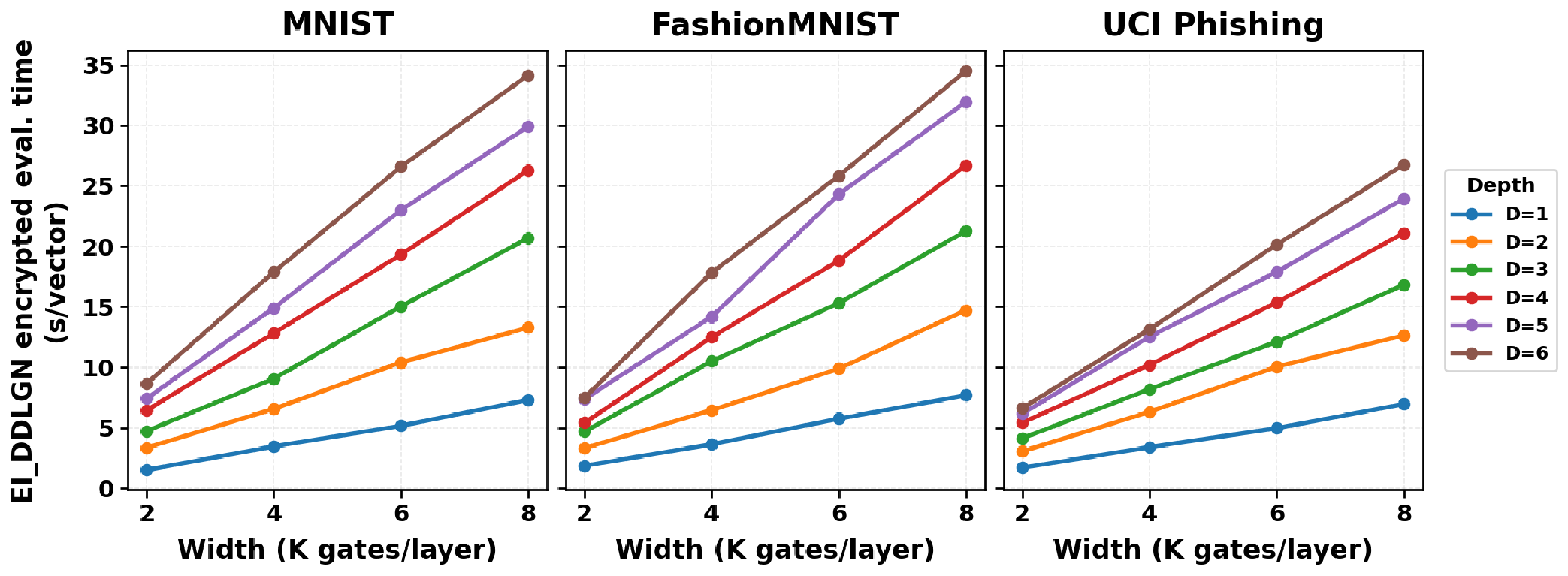}
    \caption{EI-DDLGN encrypted evaluation time as a function of width, grouped by depth and dataset.}
    \label{fig:fps_ei_encrypted_time_vs_width}
\end{figure}

In this experiment, we verify whether measured encrypted evaluation time follows the analysis from Section~\ref{sec:encrypted-inference-time}. Figure~\ref{fig:fps_ei_encrypted_time_vs_width} plots
\begin{equation}
\text{\(x\)-axis: } \frac{w}{1000},
\qquad
\text{\(y\)-axis: } T_{\mathrm{eval}}(d,w),
\end{equation}
where \(T_{\mathrm{eval}}(d,w)\) is the measured EI-DDLGN encrypted evaluation time for a model with depth \(d\) and width \(w\).

As shown above, the behavior is consistent with Eq.~(\ref{eq:eval-time-layerwise}).
For fixed depth, increasing the width increases the number of gates
available in each layer, while for fixed width, increasing the depth
increases the number of sequential layer evaluations. Consequently,
both dimensions contribute to encrypted execution cost. The observed
trends support the latency model of Section~\ref{sec:encrypted-inference-time}
and indicate that the executed PBS count provides a useful proxy for
encrypted evaluation time.

\subsection{Comparison Against Arithmetic Based FCNN Baselines}

 The goal of this experiment is to compare the proposed 
 EI-DDLGN against arithmetic based TFHE-compatible FCNN baselines 
 and assess whether the DDLGN formulation leads to a better accuracy--latency trade-off.

 For comparison, we reproduced FCNN results from \cite{stoian2023deep} 
 under the narrow-range setting with 2-bit quantization, 
 as summarized in Table~\ref{tab:sparsity_summary}. These results are averaged over 10 independently trained models using the same hyperparameters but different random seeds.
 All baseline configurations use the same three-layer fully connected architecture 
 with 192, 192, and 10 neurons from~\cite{stoian2023deep}. The FCNN variants differ in sparsity level, 
 which determines the number of active connections and, indirectly, 
 the resulting circuit bit-width and encrypted inference cost. 
\begin{table}[htpb!]
\centering
\caption{Compact MNIST comparison between EI-DDLGN and reproduced QAT-FCNN arithmetic TFHE baselines. Results are averaged over 10 independent seeds.}
\label{tab:sparsity_summary}
\setlength{\tabcolsep}{5pt}
\resizebox{\linewidth}{!}{%
\begin{tabular}{llrrl}
\toprule
Model & Size / active conn. & Acc. (\%) & Time/img (s) & Circuit bit-width \\
\midrule
EI-DDLGN Small  & 2 layers, 4K width  & 91.55 & 6.58   & N/A \\
EI-DDLGN Medium & 4 layers, 6K width  & 95.87 & 19.33  & N/A \\
EI-DDLGN Large  & 6 layers, 8K width  & 97.20 & 34.14  & N/A \\
QAT-FCNN-4      & 56 active conn.     & 91.54 & 88.24  & 6-bit: 50\%, 7-bit: 50\% \\
QAT-FCNN-6      & 84 active conn.     & 92.25 & 126.97 & 6-bit: 10\%, 7-bit: 90\% \\
QAT-FCNN-8      & 112 active conn.    & 92.45 & 136.68 & 7-bit: 100\% \\
QAT-FCNN-10     & 140 active conn.    & 92.29 & 135.98 & 7-bit: 100\% \\
QAT-FCNN-11     & 154 active conn.    & 92.58 & 132.25 & 7-bit: 100\% \\
QAT-FCNN-12     & 168 active conn.    & 92.42 & 207.98 & 7-bit: 90\%, 8-bit: 10\% \\
\bottomrule
\end{tabular}
}
\\[2pt]
\scriptsize
\raggedright
The QAT-FCNN rows follow the 3-layer FCNN architecture reported in~\cite{stoian2023deep}: three fully connected layers with 192, 192, and 10 neurons. The circuit bit-width column shows the distribution of the maximum encrypted circuit bit-width across the 10 independent runs for each QAT-FCNN sparsity setting.
\end{table}

 The QAT-FCNN baselines achieve accuracies between 91.54\% and 92.58\%, 
 with inference times ranging from 88.24\,s to 207.98\,s per image depending on 
 sparsity level and resulting circuit bit-width. 
 By contrast, the EI-DDLGN Small model is essentially accuracy-matched with 
 QAT-FCNN-4, reaching 91.55\% accuracy in 6.58 seconds per image compared 
 with 88.24 seconds per image, achieving a \(13.4\times\) latency improvement. 
  EI-DDLGN Medium reaches 95.87\% accuracy, 
 exceeding the best QAT-FCNN accuracy by 3.29\%, 
 while achieving a \(6.8\times\) latency improvement over the best-accuracy QAT-FCNN row. 
  EI-DDLGN Large reaches 97.20\% accuracy, improving over the best QAT-FCNN accuracy by 
 4.62\%, while still achieving a \(3.9\times\) 
 latency improvement over that baseline.

Another important observation from Table~\ref{tab:sparsity_summary} 
is that QAT-FCNN performance remains strongly coupled to circuit bit-width. 
Most baseline configurations operate predominantly at 7-bit width, 
while the highest active-connections configuration occasionally reaches 8-bit width 
and exhibits the highest latency. 
In contrast, the EI-DDLGN evaluates Boolean gates directly under TFHE 
and does not rely on quantized integer accumulation in the hidden layers. 
This difference leads to a more predictable and structurally aligned encrypted 
execution model, and explains why EI-DDLGN provides a substantially better 
accuracy--latency trade-off in the evaluated setting.

An important observation is that the latency improvements reported in
Table~\ref{tab:sparsity_summary} cannot be explained solely by
MFW-PBS Bypass. The measured speedups are substantially larger than
the PBS bypass rates reported in Section~\ref{sec:pbs-analysis},
indicating that the primary efficiency gain originates from the
Boolean-native structure of DDLGNs themselves. By replacing arithmetic
accumulation with learned Boolean computations, EI-DDLGN changes the
underlying encrypted computation model, while MFW-PBS Bypass provides
an additional optimization on top of that architectural advantage.

\section{Conclusion}
\label{Sec:Conclusion}

This paper presented EI-DDLGN, the first in-depth study of Deep
Differentiable Logic Gate Networks (DDLGNs) under TFHE. By expressing
inference as the homomorphic evaluation of learned Boolean functions,
EI-DDLGN provides a Boolean-native alternative to arithmetic neural network
inference.

We characterized the PBS cost structure of TFHE-based DDLGN inference,
introduced Model-Fixed-Wire PBS Bypass (MFW-PBS Bypass), and showed
that encrypted execution depends jointly on model size, learned
Boolean-function distribution, and propagated wire status.

Experiments across 72 depth--width configurations demonstrated clear
accuracy--latency and accuracy--PBS budget trade-offs and achieved up
to a \(13.4\times\) latency reduction over an accuracy-matched
QAT-FCNN baseline. These results suggest that Boolean-native neural
architectures constitute a promising direction for efficient
privacy-preserving inference under TFHE.

\bibliographystyle{splncs04}
\bibliography{references}
\end{document}